\documentclass[aps,prl,twocolumn,superscriptaddress,showpacs,nobibnotes]{revtex4}
\usepackage{graphicx}
\usepackage{amsmath}
\usepackage{epstopdf}
\usepackage{color}

\usepackage[english]{babel}
\usepackage[applemac]{inputenc}
\usepackage{pdfpages}

\usepackage{hyperref}
\hypersetup{
    colorlinks,
    citecolor=green,
    filecolor=black,
    linkcolor=blue,
    urlcolor=red
}

\begin{document}

\title{Conductivity Corrections for Topological Insulators with Spin-Orbit Impurities: \\
A New Hikami-Larkin-Nagaoka Formula}
\date{\today}
\author{P. Adroguer}\thanks{The two first authors contributed equally.}
\affiliation{Institute for Theoretical Physics and Astrophysics, W\"urzburg University, Am Hubland, 97074 W\"urzburg, Germany}

 \author{Weizhe E. Liu}\thanks{The two first authors contributed equally.}
  \affiliation{School of Physics, The University of New South Wales, Sydney 2052, Australia}
 
 \author{D. Culcer}
\affiliation{School of Physics, The University of New South Wales, Sydney 2052, Australia}

\author{E. M. Hankiewicz}\email[Author to whom correspondence should be addressed : ]{hankiewicz@physik.uni-wuerzburg.de}
\affiliation{Institute for Theoretical Physics and Astrophysics, W\"urzburg University, Am Hubland, 97074 W\"urzburg, Germany}

\begin{abstract}
The Hikami-Larkin-Nagaoka (HLN) formula [Prog. Theor. Phys. \textbf{63}, 707 (1980)] describes the quantum corrections to the magnetoconductivity of a quasi-2D electron gas (quasi-2DEG) with parabolic dispersion. It predicts a crossover from weak localization to antilocalization as a function of the strength of scattering off spin-orbit impurities. Here, we derive the conductivity correction for massless Dirac fermions in 3D topological insulators (3DTIs) in the presence of spin-orbit impurities. We show that this correction is always positive and therefore we predict weak antilocalization for every value of the spin-orbit disorder. Furthermore, the correction to the diffusion constant is surprisingly linear in the strength of the impurity spin-orbit. Our results call for a reinterpretation of experimental fits for the magnetoconductivity of 3D TIs which have so far used the standard HLN formula.  
\end{abstract}

\maketitle

 \textit{Introduction.} The problem of the diffusion of the surface states of 3DTIs is a complex one due to a variety of competing phenomena. The most striking is the fact that these surface states are described by a Dirac Hamiltonian \cite{Fu:2007,Chen:2009}, which gives rise to weak antilocalization (WAL) in the presence of scalar disorder \cite{McCann:2006,Tkachov:2011,Adroguer:2012}. The WAL correction can be affected by the interaction of the surface states with the residual bulk states \cite{Garate:2012}, the thickness of the film \cite{Shan:2012}, or electron-electron interactions \cite{Koenig:2013,Lu:2014} changing its sign and turning it into weak localization (WL). At the same time, since 3DTIs have strong spin-orbit coupling, one expects spin-orbit coupled impurities to have a strong effect on transport, different however than in graphene where two valleys are present \cite{Kechedzy:2007,McCann:2012}. 
Surprisingly this problem has received virtually no attention \cite{Culcer_TI_AHE_PRB11}  and so far the Dirac nature of the states, that manifest itself in the angular dependence of the Green functions, has not been taken into account in studies of spin-orbit impurities \cite{Shan:2012}. 
The problem is even more complicated in a transverse magnetic field.
 
\begin{figure}
\centering
\includegraphics[width= 0.45 \textwidth]{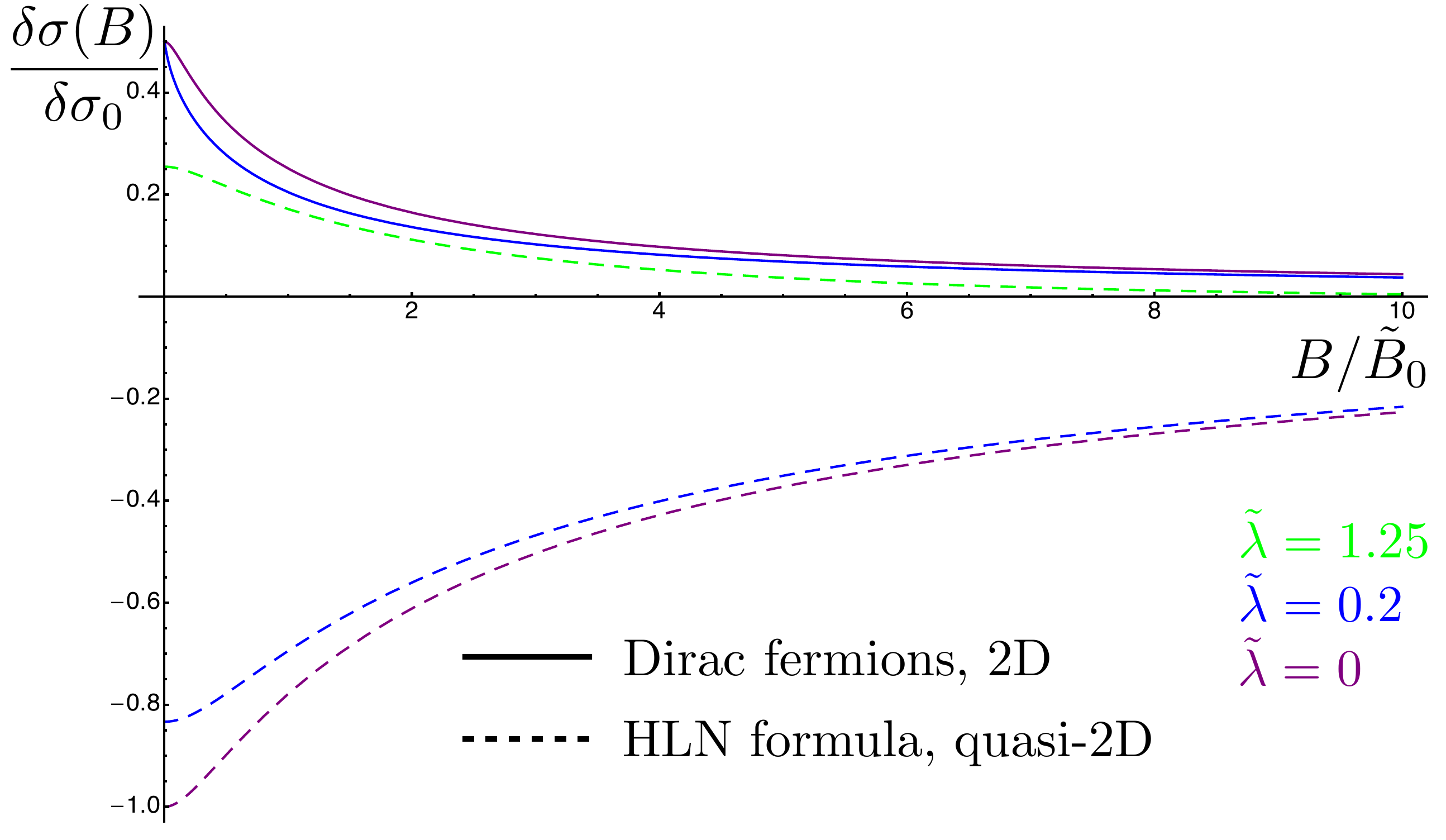}
\caption{Comparison of the quantum correction to the conductivity as a function of the magnetic field for massless Dirac fermions (solid lines) and the Hikami-Larkin-Nagaoka formula for a quasi-2D system (dashed lines) for different values of the concentration of the spin-orbit impurities. Purple plots are for purely scalar disorder ($\tilde{\lambda} =0$), blue (green) plots for $\tilde{\lambda} = 0.2$ ($\tilde{\lambda} = 1.25$). The magnetic field is renormalized by $\tilde{B}_0 = e / (2 \pi \hbar v_\text{F}^2 \tau_0^2)$, where $\tau_0$ is the elastic scattering time in the absence of spin-orbit impurities and $v_\text{F}$ is the Fermi velocity.}
\label{fig:Qcorrec_HLN3D_alphaB}
\end{figure}

The formula commonly used to fit the magnetoconductance experiments on 3DTIs \cite{Checkelsky:2011,Zhang:2013,Lang:2013,Kim:2013,Bao:2013,Lin:2013,Chen:2011,Bardarson:2013} was derived by Hikami, Larkin and Nagaoka (HLN) \cite{Hikami:1980}. The HLN formula, however, lacks important features relevant to 3DTIs: it is derived for quasi-2DEGs with a parabolic electron dispersion (impurities are treated as three-dimensional objects), so it accounts neither for the Dirac nature of the surface states nor for their strictly two-dimensional character. 
Fig.~\ref{fig:Qcorrec_HLN3D_alphaB} is the main result of this paper, and shows the different effects of spin-orbit scattering for the HLN formula, namely WL to WAL crossover with the strength of spin-orbit impurity scattering, and for Dirac fermions, where WAL appears regardless of the strength of spin-orbit impurities. We only observe a convergence of the two formulas for large values of the strength of the impurity spin-orbit scattering (when both equations are no longer valid)\footnote{For 3DTIs, the HLN formula predicts weak antilocalization for $\tilde{\lambda}$ close to 1, see Supplementary material for more details.}. Moreover, the HLN formula gives the wrong value for the diffusion constant when the impurity spin-orbit coupling vanishes: it does not capture the fact that the absence of backscattering for Dirac fermions doubles the diffusion constant as compared to conventional electrons.

In this paper, we present the full calculation of coherent diffusive transport of Dirac fermions in the presence of both scalar and spin-orbit coupled impurities. We first show that, because of the winding of the spin around the Fermi surface, the Dirac nature of the surface states breaks the mirror symmetry around the $xy$-plane (the disorder potential no longer commutes with the kinetic Hamiltonian), and thus allows for a correction to physical quantities, such as the classical conductivity and the diffusion constant, which is \textit{linear} in the strength of the disorder spin-orbit coupling as opposed to the quadratic dependence observed for a parabolic dispersion. Moreover, we show that WAL appears for any strength of the spin-orbit coupled disorder for massless Dirac fermions, as opposed to the case of electrons with parabolic dispersion, where a crossover from WL to WAL (no correction) is observed for quasi-2DEGs (strictly 2DEGs).


\smallskip
\textit{Model.} The Dirac Hamiltonian for the surface states of 3DTIs, including a random uncorrelated Gaussian disorder potential comprising both scalar and spin-orbit impurities, with impurity concentration $n_\text{i}$, reads:
\begin{equation}
\mathcal{H} = \hbar v_\text{F} (k_x \sigma^y - k_y \sigma^x) + V(\vec{k},\vec{k'}) ,
\label{eq:ham_kin}
\end{equation}
with $V(\vec{k},\vec{k'}) = U \displaystyle\, p_{{\vec{k}}{\vec{k}'}} \left[ \openone +  i \lambda (\vec{k} \times \vec{k'}) \cdot \vec{\sigma}\right]$ the disorder potential. Here $\openone$ is the $2 \times 2$ identity matrix in spin space, $p_{{\vec{k}}{\vec{k}'}} = \sum_\text{I} e^{-i({\vec{k}} - {\vec{k}'})\cdot{\vec{R}}_\text{I}}$ with `I' labelling the random locations of the impurities, and we have assumed a short-range impurity potential so that $U$ is not a function of wave vector. The disorder correlator $b(\theta,\theta')$ is
\begin{equation}
 \Big\langle V(\vec{k},\vec{k'})  V(\vec{k'},\vec{k}) \Big\rangle = \gamma_0 \Big(\text{I}_0+i \tilde{\lambda} \text{I}_1\sin \gamma + \tilde{\lambda}^2 \text{I}_2\sin^2\gamma \Big)\!,
\end{equation}
where $\gamma=\theta'-\theta$ and $\theta$ ($\theta'$) is the polar angle of the wave vector $\vec{k}$ ($\vec{k}'$). We have introduced $\gamma_0 = n_\text{i} U^2$ to quantify the disorder strength, and the dimensionless parameter $\tilde{\lambda} = \lambda k_\text{F}^2$ to describe the relative strength of the scalar and spin-orbit coupled disorder. We work in the limit $\tilde{\lambda} \ll 1$, meaning that the spin-orbit scattering length is much larger than the mean-free path. The $\text{I}_i$ are: $\text{I}_0=\openone \otimes \openone$, $\text{I}_1=\sigma^z \otimes \openone - \openone \otimes \sigma^z$, and $\text{I}_2=\sigma^z \otimes \sigma^z$. The Hamiltonian~(\ref{eq:ham_kin}) preserves time-reversal symmetry (TRS). Due to the non-commutativity of the Pauli matrices in the band and impurity Hamiltonians, terms linear in $\tilde{\lambda}$ affect charge and spin dynamics, as opposed to the case of spinless electrons where these linear terms are absent.

The bare Green function reads:
\begin{equation}
G^0_{ss'}(\vec{k}) = \frac{1}{2} \bigg(\frac{ \openone_{ss'}+ \cos \theta \sigma_{ss'}^y - \sin \theta \sigma_{ss'}^x }{E - \hbar v_\text{F} k \pm i 0^+}\bigg).
\end{equation}
The lifetime of a particle with wave vector $\vec{k}$ in the weak disorder limit is the imaginary part of the self energy $\Sigma$:
\begin{equation}
\Sigma(\vec{k}) = \int \frac{d \vec{k'}}{(2 \pi)^2} b(\theta,\theta') G^0(\vec{k'}) \; .
\end{equation}
We introduce the two characteristic times $\tau$ and $\tau^*$ as $\frac{\hbar}{\tau} = \pi \rho(E_\text{F}) \gamma_0 (1 + \frac{\tilde{\lambda}^2}{2})$ and $\frac{\hbar}{ \tau^*} = \pi \rho(E_\text{F}) \gamma_0 \tilde{\lambda} $ where $\rho(E_\text{F}) = \frac{E_\text{F}}{2 \pi \hbar^2 v_\text{F}^2}$ is the density of states at the Fermi energy. We obtain for the imaginary part of the self energy: $\displaystyle - \text{Im} \Sigma(\vec{k}) =  \frac{\hbar}{2 \tau} \openone + \frac{\hbar}{2 \tau^*} (\cos \theta \sigma^y - \sin \theta \sigma^x)$. Near the Fermi surface the retarded and advanced Green functions  $G^\text{R/A} = [E - (H_0 \pm \Sigma)]^{-1}$ take the form
 \begin{equation}
G^\text{R/A} (\vec{k}) = \frac{1}{2} \bigg(\frac{ \openone +  \cos \theta \sigma^y - \sin \theta  \sigma^x}{ E - \hbar v_\text{F} k \pm \frac{i\hbar}{2 \tau_\text{e}}} \bigg),
\end{equation}
where the corresponding mean free-time between two scattering events $\tau_\text{e}$, derived through the Fermi golden rule, obeys the Matthiessen rule $\frac{1}{\tau_\text{e}} = \frac {1 }{\tau} + \frac{1}{ \tau^*} = \frac{ \pi \rho(E_\text{F}) \gamma_0}{\hbar} \left(1 + \tilde{\lambda} + \frac{\tilde{\lambda}^2}{2} \right)$. Due to the spin structure of both the Green function and the scattering potential, this elastic mean free-time shows an unusual linear dependence in the spin-orbit scattering strength $\tilde{\lambda}$, as opposed to the case of non-relativistic electrons \cite{Hikami:1980}. The mean-free time is no longer an even function of $\tilde{\lambda}$, since the winding of the spin around the Fermi surface for Dirac fermions defines unequivocally the direction of the $z$-axis.

\smallskip
\textit{Diffuson and renormalized current operator.} The ladder diagrams are responsible for the difference between the elastic scattering time $\tau_\text{e}$ and the transport time $\tau_\text{tr}$ appearing in the diffusion constant $D = \frac{v_\text{F}^2 \tau_\text{tr}}{d}$ in a $d$-dimensional material. For Dirac fermions (graphene, 3DTI surface states), in the presence of point-like scalar disorder scattering, a doubling of the transport time $\tau_\text{tr}= 2 \tau_\text{e}$ is observed. The diffuson structure factor $\Gamma^\text{D}$ obeys the Bethe-Salpeter equation (see Fig.~2)~:
\begin{align}
&\Gamma_{\substack{\alpha \beta \\ \gamma \delta}}^D(\theta,\theta',\vec{q}) = b_{\substack{\alpha \beta \\ \gamma \delta}}(\theta,\theta') \ +  \\
& \int \frac{d \vec{k''}}{(2 \pi)^2} \Gamma_{\substack{\alpha \mu \\ \gamma \lambda}}^D(\theta,\theta'',\vec{q})  G_{\mu \nu}^R(\vec{k''}) G_{\kappa \lambda}^A(\vec{k''}-\vec{q}) b_{\substack{\nu \beta \\ \kappa \delta}}(\theta'',\theta') \nonumber \; ,
\end{align}
in which $\theta$ ($\theta'$) denotes the direction of the incoming (outgoing) wave vector and Greek symbols are the spin indices. We will drop from now on the spin indices to simplify the equations.

\begin{figure}
\centering
\includegraphics[width= 0.45 \textwidth]{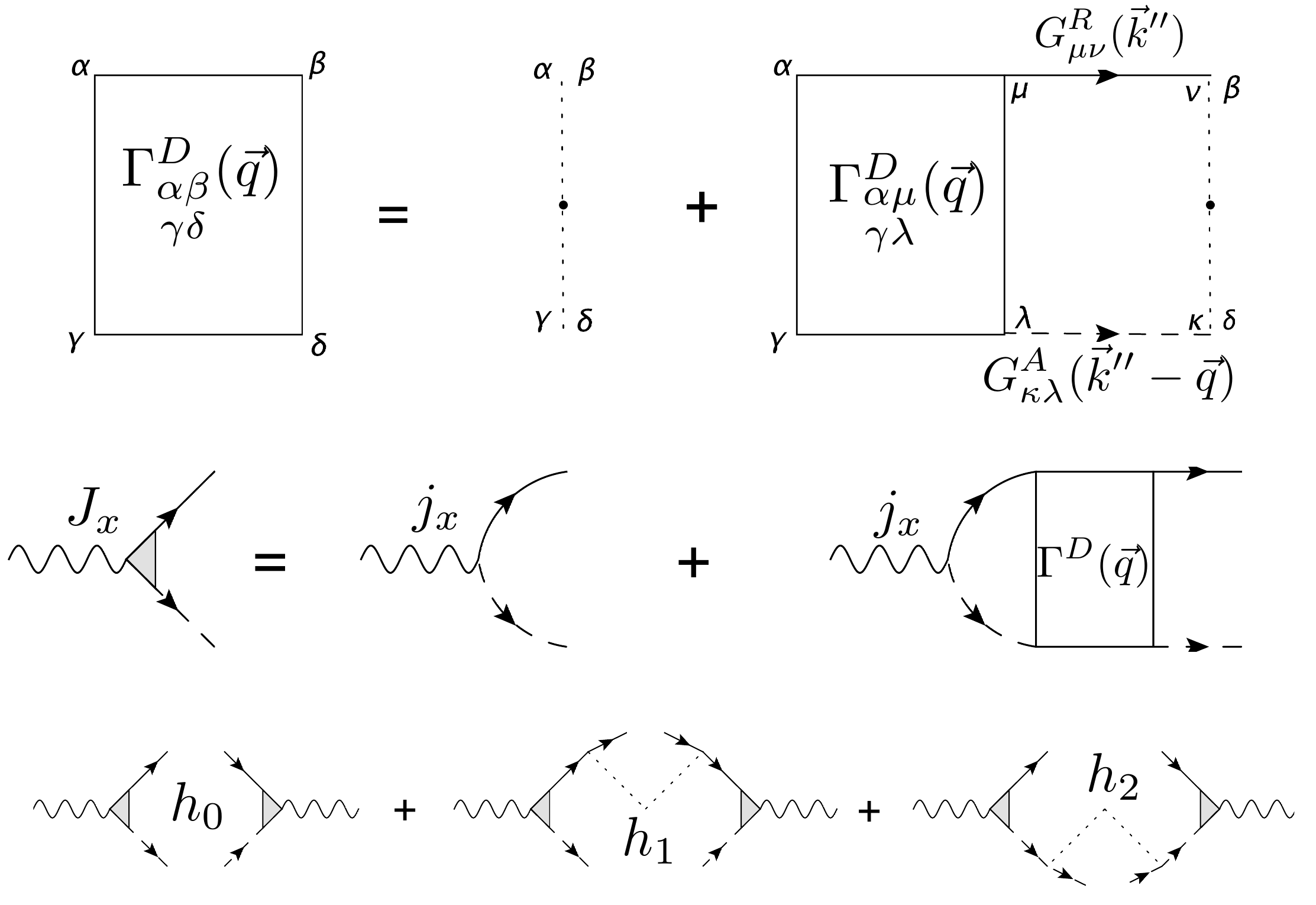}
\caption{Diagrammatic representation of the calculations. The first row represents the Bethe-Salpeter equation obeyed by the diffuson structure factor where the Greek symbols describe spin indices. The second row shows the current renormalization. 
The third row depicts the three Hikami boxes needed for the WAL calculation.}
\label{fig:diagrams}
\end{figure}

The Fourier decomposition of this diffuson structure factor as $\Gamma^\text{D}(\theta,\theta',\vec{q}) = \sum_{n,m} \Gamma^\text{D}_{n,m}(\vec{q}) e^{i(m \theta'-n \theta)}$ allows a solution of the equation perturbatively in $\tilde{\lambda}$. Up to second order, this diffuson structure factor acquires a non-trivial structure
 with angular components (e.g $\Gamma^\text{D}_{0,1} \neq 0 $) that does not appear with isotropic scattering, even for Dirac fermions. This is due to the presence of higher harmonics in both the disorder correlator $b(\theta,\theta')$, and the Green functions  (see supplementary material for details).

The diffuson structure factor renormalizes the current operators $j(\vec{k})$ (cf. Fig.~\ref{fig:diagrams}) as:
\begin{equation}
 J(\vec{k}') = j(\vec{k}') + \int_{\vec{k}} G^A(\vec{k}) j(\vec{k}) G^R(\vec{k}) \Gamma^\text{D} (\theta, \theta',\vec{0}) \; .
 \end{equation}
We find explicitly for the renormalized current operator along the $x$-direction
  \begin{equation}
 J_x (\theta') = - e v_\text{F} \big[ 2 (1 - \tilde{\lambda} + \frac{9 \tilde{\lambda}^2}{8}) \sigma^y - \frac{i\tilde{\lambda}^2e^{- 2 i \theta'} }{8} \sigma^+ + h.c. \big],
 \end{equation}
where $h.c.$ denotes the Hermitian conjugate and $\sigma^\pm = \sigma^x \pm i \sigma^y$. For pure scalar disorder $\tilde{\lambda} \to 0$ we recover the doubling of the current operator $J_x = 2 j_x$. Once again, the spin structure of the band and disorder Hamiltonians is responsible for a non-trivial angular dependence of this renormalized current operator, no longer proportional to the original current operator $j_x = - e v_\text{F} \sigma^y$.


\smallskip
\textit{Longitudinal conductivity and diffusion constant.} It is now possible to compute the longitudinal conductivity through the Kubo formula as~:
\begin{equation}
\sigma^\text{Dr}_{xx}  = \frac{\hbar}{2\pi \Omega} {\rm Re} \,\text{Tr} \left[J_x(\vec{k}) G^R(\vec{k}) j_x(\vec{k}) G^A(\vec{k}) \right],
\end{equation}
where $\text{Tr}$ denotes a sum over both spins and wave vector $\vec{k}$, and $\Omega$ is the volume. Up to second order in $\tilde{\lambda}$, we find~:
\begin{equation}
\sigma^\text{Dr}_{xx} = e^2 \rho(E_\text{F}) v_\text{F}^2 \tau_\text{e} \bigg( 1 - \tilde{\lambda} + \frac{5 \tilde{\lambda}^2}{4} \bigg).
\end{equation} 
The diffusion constant $D = v_\text{F}^2 \tau_\text{e} \left( 1 - \tilde{\lambda} + \frac{5 \tilde{\lambda}^2}{4} \right)$ shows a linear dependence on the spin-orbit scattering strength, as distinct from the case of non relativistic electrons. As this diffusion constant is a crucial parameter in weak antilocalization, we expect the behavior of Dirac fermions to be different from the usual HLN formula \cite{Hikami:1980}.


\smallskip
\textit{Weak antilocalization correction.} The weak antilocalization correction requires the three different contributions pictured in Fig.~\ref{fig:diagrams}, as is usual for Dirac fermions \cite{McCann:2006}. The Cooperon structure factor $\Gamma^\text{C}$ (maximally crossed diagrams) accounts for the quantum interferences of closed paths during diffusion. Because of TRS, it can be obtained from the diffuson structure factor by twisting the retarded or advanced branch. It obeys:
\begin{align}
&\Gamma_{\substack{\alpha \beta \\ \gamma \delta}}^C(\theta,\theta',\vec{q}) = b^C_{\substack{\alpha \beta \\ \gamma \delta}}(\theta,\theta') \ +  \\
& \int \frac{d \vec{k''}}{(2 \pi)^2} \Gamma_{\substack{\alpha \mu \\ \gamma \lambda}}^C(\theta,\theta'',\vec{q})  G_{\mu \nu}^R(\vec{k''}) G_{\lambda \kappa}^A(\vec{q} - \vec{k''}) b^C_{\substack{\nu \beta \\ \kappa \delta}}(\theta'',\theta') \nonumber \; ,
\end{align}
where $b^C= \gamma_0 \Big(\text{I}_0+i \tilde{\lambda} \tilde{\text{I}}_1\sin \gamma - \tilde{\lambda}^2 \text{I}_2\sin^2\gamma \Big)$ and $\tilde{\text{I}}_1=\sigma^z \otimes \openone + \openone \otimes \sigma^z$.
We also expand $\Gamma^\text{C}(\theta,\theta') = \sum_{n,m} \Gamma^\text{C}_{n,m} e^{i(m \theta'-n \theta)}$, keeping only the 9 terms in $1/q^2$ up to second order in  $\tilde{\lambda}$ (the terms with $n,m \in {0, \pm1}$) (see supplementary material).


\smallskip
 It is then possible to calculate the three Hikami boxes pictured in Fig.~\ref{fig:diagrams} for each of these nine terms (full expressions in supplement). The contribution of each $\Gamma^\text{C}_{n,m}$ mode can be collected in three different groups, depending on the value $i = \vert n \vert + \vert m \vert =  0, 1,2 $. The respective weights $w_i$ of these contributions are $w_0 = \frac{1}{2} - \tilde{\lambda} + \frac{3 \tilde{\lambda}^2}{2}$, $w_1 = \frac{\tilde{\lambda}}{4} - \frac{\tilde{\lambda}^2}{2} $ and $w_2 = \frac{\tilde{\lambda}^2}{8}$. Summing these 9 contributions, we obtain the WAL correction~:

\begin{equation}
\delta \sigma_\text{D.F} = \frac{e^2}{2 \pi h} \ln \left( \frac{\tau_{\phi}}{\tau_\text{e}} \right) \label{eq:delta1} \; .
\end{equation}

This expression should be compared with the formula for non-relativistic electrons for strictly 2D and quasi-2D systems~:
\begin{equation}
\arraycolsep 0.3ex
\begin{array}{rl}
\displaystyle \delta \sigma_\text{HLN}^\text{(2D)} = & \displaystyle - \frac{e^2}{\pi h} \ln \left( \frac{1 + \tilde{\lambda}^2/2}{ \frac{\tau_\text{e}}{ \tau_{\phi}} + \tilde{\lambda}^2/2} \right) \; ,\\ [3ex]
\displaystyle \delta \sigma_\text{HLN}^\text{(q-2D)} = & \displaystyle - \frac{e^2}{\pi h} \Bigg[ \ln \bigg( \frac{1 + \frac{2 \tilde{\lambda}^2}{3}}{ \frac{\tau_\text{e}}{ \tau_{\phi}}  + \frac{2 \tilde{\lambda}^2}{3}} \bigg) - \frac{1}{2}  \ln \bigg( 1+ \frac{8 \tilde{\lambda}^2}{9}\frac{\tau_{\phi}}{\tau_\text{e}}\bigg) \Bigg].
\end{array}
\end{equation}

\begin{figure}
\centering
\includegraphics[width= 0.50 \textwidth]{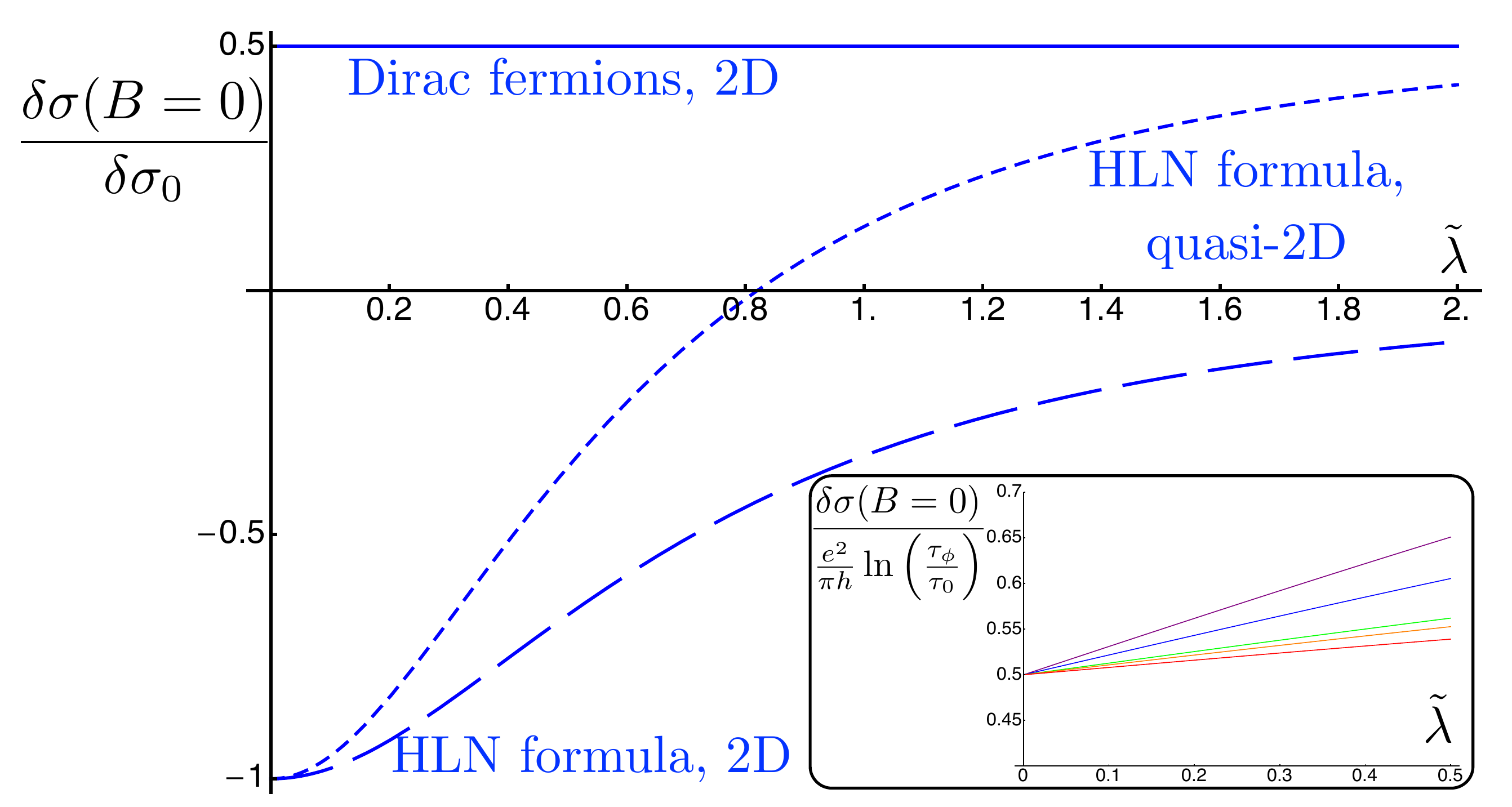}
\caption{Comparison of the quantum correction to conductivity as a function of the spin-orbit scattering strength $\tilde{\lambda}$ for massless Dirac fermions (solid line),  Hikami-Larkin-Nagaoka formula for strictly 2D system (dashed line) and quasi-2D system (dotted line). The conductivities are given in units of $\delta \sigma_0 =  \frac{e^2}{ \pi h} \ln \left( \frac{\tau_{\phi}}{\tau_\text{e}} \right)$. To emphasize the dependence of WAL 
on the spin-orbit impurity strength for massless Dirac fermions, the inset shows WAL in units independent of $\tilde{\lambda}$ i.e. renormalized by $\frac{e^2}{ \pi h} \ln \left( \frac{\tau_{\phi}}{\tau_{0}} \right)$ ($\tau_0$ is the scattering time in the absence of spin-orbit impurities).
The ratio $\tau_\phi / \tau_0$ varies over two decades from 5 (purple), 10 (blue), 50 (green), 100 (orange) to 500 (red).}
\label{fig:Qcorrec_alpha}
\end{figure}

These three formulas are plotted in Fig.~\ref{fig:Qcorrec_alpha} as a function of $\tilde{\lambda}$. We have renormalized these corrections by $\delta \sigma_0 =  \frac{e^2}{\pi h} \ln \left( \frac{\tau_{\phi}}{\tau_\text{e}} \right)$, where $\tau_\text{e}$ depends on the model and is a function of $\tilde{\lambda}$. We have set the ratio $\frac{\tau_{\phi}}{\tau_0}$ where $\tau_0$ is the value of $\tau_\text{e}$ in the absence of spin-orbit scattering $\tau_0 = \frac{\hbar}{\pi \rho (E_\text{F}) \gamma_0}$ to be equal to $10$ in agreement with what is measured experimentally\cite{Muhlbauer:2014,Chiu:2013,Zhao:2013}. We observe that the Dirac fermions remain in the same symmetry class (symplectic, with WAL), whereas the HLN formula shows a crossover from the orthogonal symmetry class (WL) to either no correction for strictly 2DEG, or WAL for quasi-2DEG.

Due to the renormalization of the scattering time by spin-orbit impurities, one could also interpret Eq.~(\ref{eq:delta1}), as increasing with $\tilde{\lambda}$ if one renormalizes conductivity corrections by the scattering time in the absence of spin-orbit impurities $\tau_0$ (namely normalized by $\ln \tau_{\phi}/\tau_{0}$) : 
\begin{equation}
\delta \sigma_\text{D.F} = \frac{e^2}{2 \pi h}\left[ \ln \left( \frac{\tau_{\phi}}{\tau_{0}} \right) + \tilde{\lambda} + O(\tilde{\lambda}^2) \right].
\end{equation}
The inset to Fig.~3 shows the linear dependence of the normalized correction to the conductivity as a function of $\tilde{\lambda}$ for massless Dirac fermions. As $\tilde{\lambda} = \lambda k_\text{F}^2$, one can experimentally probe this linear dependence by varying the Fermi wave vector using an electrostatic gate, for a constant strength of spin-orbit impurity scattering. 


\smallskip
\textit{WAL correction as a function of the magnetic field.} The well-known HLN formula \cite{Hikami:1980} describes the quantum correction to conductivity as a function of an applied magnetic field for non-relativistic electrons in presence of both scalar and spin-orbit impurities, where the only relevant parameter is the diffusion constant. In contrast, in our problem, each mode $\Gamma^\text{C}_{n,m}$ obeys a diffusion equation with a diffusion constant $D_i$ depending on $i = \vert n \vert + \vert m \vert$. Namely, these diffusion constants are $D_0 = v_\text{F}^2 \tau_\text{e}/(1 + \tilde{\lambda}^2/2)$, $D_1 = 2 v_\text{F}^2 \tau_\text{e}/\tilde{\lambda} $ and $D_2 = 2 v_\text{F}^2 \tau_\text{e}/ \tilde{\lambda}^2$.

Inserting the magnetic field through the Peierls substitution maps this diffusion equation onto a Schr\"odinger equation with the equivalence $D_i \leftrightarrow \hbar/2m$ and $e \leftrightarrow 2e$ \cite{Altshuler:1981,Altshuler:1980,Montambaux:book}. It follows that the introduction of the magnetic field reduces the contribution of each mode to the WAL correction by a factor~:
\begin{equation}
g_i(B) = \frac{ \Psi \left( \frac{1}{2} + \frac{B^\text{e}_i}{B} \right) - \Psi \left( \frac{1}{2} + \frac{B^\phi_i}{B} \right) }{\ln \frac{\tau_\phi}{\tau_\text{e}}} \; ,
\end{equation}
where we define the characteristic fields $B^\text{e}_i = \frac{\hbar}{4 D_i \tau_\text{e}}$ and $B^\phi_i = \frac{\hbar}{4 D_i \tau_\phi}$. 
Collecting all the modes we obtain~:
\begin{equation}
\delta \sigma_{D.F.} (B) = \frac{e^2}{ \pi h} \ln \frac{\tau_\phi}{\tau_\text{e}} \sum_{i=1}^3 a_i w_i g_i (B)\; , \label{eq:results}
\end{equation}
where the mode weights $a_i$ are $a_1=1$, and $a_2=a_3=4$, with the $w_i$s defined before Eq.~(\ref{eq:delta1}).
The magnetoconductivity corrections for massless Dirac fermions are presented in Fig.~1.

 
\smallskip
\textit{Discussion.} Our results show many differences from the HLN formula widely used to fit magnetotransport experiments of 3DTI surface states. The first is that the winding of the spin around the Fermi surface breaks the mirror symmetry around the $xy$-plane, so it is now possible to obtain a linear dependence of measurable quantities on $\tilde{\lambda}$, the strength of the spin-orbit disorder. Such a linear dependence is observed in the mean-free time $\tau_\text{e}$, the longitudinal Drude conductivity $\sigma_{xx}$, and the diffusion constant $D$.

A second type of difference emerges from the anisotropy of the Green functions for massless Dirac model. This anisotropy together with the anisotropy coming from the spin-orbit impurities leads to 9 different Fourier modes in the Cooperon to the second order in  $\tilde{\lambda}$ as opposed to only one mode for the HLN model. Moreover, our expansion in the spin-orbit impurity strength shows more explicitly the fact that this calculation is perturbative in $\tilde{\lambda}$ and should be restricted to small values of the perturbative parameter as the odd powers of the series expansion contribute negatively to the conductivity.

In general symmetry terms, the massless Dirac fermion model stays in the symplectic class for all values of the impurity spin-orbit coupling, as the square of the time reversal operator $\Theta^2 = - \openone$. This explains why WAL is always observed for Dirac fermions. In the HLN formula, the introduction of the impurity spin-orbit coupling is responsible for a crossover from the orthogonal class $\Theta^2 = \openone$ when $\tilde{\lambda}=0$ to the symplectic ("pseudo-unitary"\footnote{Although TRS is preserved, the spin-orbit coupling only affects the $z$-component of the spin for a strictly 2D system, and the triplet state with no net magnetization along the $z$-axis is not suppressed. As a consequence, the singlet and triplet compensate each other, resulting in no correction to conductivity. This is similar to the unitary class where TRS is broken and all four states are suppressed.}) class when $\tilde{\lambda} \to \infty$ for quasi-2DEGs (strictly 2DEGs).


\smallskip
\textit{Conclusions.} We have derived the magnetoconductivity corrections for the surface states of 3D TI in the presence of scalar and spin-orbit impurity disorder. This is expected to be directly relevant to the experimental analysis of these materials. We showed the profound difference between the HLN model for parabolic dispersion and the massless Dirac fermion model. For the latter we predict WAL in the presence as well as in the absence of spin-orbit impurity scattering, while the diffusion constant and the longitudinal conductivity are renormalized to the first order in the spin-orbit scattering strength.


\smallskip
\textit{Acknowledgements.} We acknowledge useful discussions with L. Molenkamp, T. Sch\"apers, M. Fuhrer and Y. Li. We thank the German Science Foundation (DFG), grants No HA 5893/4-1 within SPP 1666, as well the ENB graduate school "Topological insulators" for the financial support. 


\bibliographystyle{apsrev} 

\onecolumngrid
\appendix
\newpage
\begin{center}
{\textbf{\large{Supplementary material}}}
\vspace{-0.3cm}
\end{center}

\section{Solution of the Bethe-Salpeter equation for the diffuson}

The calculation of the classical conductivity requires the current operator renormalization, through the insertion of the diffuson, also known as the ladder diagram. To obtain this diffuson, we need to solve the Bethe-Salpeter equation~:
\begin{align}
&\Gamma_{\alpha \beta,\gamma \delta}^\text{D}(\theta,\theta',\vec{q}) = b_{\alpha \beta, \gamma \delta}(\theta,\theta') +  \int \frac{d \vec{k''}}{(2 \pi)^2} \Gamma_{\alpha \mu, \gamma \lambda}^\text{D}(\theta,\theta'',\vec{q})  G_{\mu \nu}^R(\vec{k''}) G_{\kappa \lambda}^A(\vec{k''}-\vec{q}) b_{\nu \beta, \kappa \delta}(\theta'',\theta') \; ,
\end{align}
where $\theta$ ($\theta'$) denotes the direction of the incoming (outgoing) wave vector and Greek symbols describe the spin indices.

Our first step is to expand all the quantities in Fourier series in order to remove the angular dependence. We write~:
\begin{eqnarray}
\Gamma_{\alpha \beta, \gamma \delta}^\text{D}(\theta,\theta',\vec{q}) &=& \sum_{n,m} \Gamma_{(n,m)\alpha \beta,\gamma \delta}^\text{D}(\vec{q}) e^{i(m \theta'-n \theta)} \; , \\
b_{\alpha \beta,\gamma \delta}(\theta,\theta') &=& \sum_{n,m} b_{(n,m)\alpha \beta,\gamma \delta} e^{i(m \theta'-n \theta)} \; ,
\end{eqnarray}
as $b$ only depends on $\theta' -\theta$, we will use $b_{(n,m)} = b_{(n)} \delta_{n,m}$ where $\delta$ is the Kronecker symbol. Looking at each mode in the Fourier expansion, we obtain a system of coupled equations~:
\begin{equation}
\Gamma^\text{D}_{(n,m)\alpha \beta,\gamma \delta} (\vec{q}) = b_{(n) \alpha \beta, \gamma \delta} \delta_{n,m} + \sum_k \Gamma^\text{D}_{(n,m+k) \alpha \mu,\gamma \lambda}(\vec{q}) P^\text{D}_{(k) \mu \nu, \lambda \kappa}(\vec{q}) b_{(m)\nu \beta,\kappa \delta} \label{eq:Gamma} \; ,
\end{equation}
where $P^\text{D}_{(k)\mu \nu, \lambda \kappa}(\vec{q})$ is defined as ~: 
\begin{eqnarray}
P^\text{D}_{(k)\mu \nu, \lambda \kappa}(\vec{q}) &=& \int \frac{d \vec{k''}}{(2 \pi)^2}  e^{ik\theta''} G_{\mu \nu}^R(\vec{k}'') G^A_{\kappa \lambda} (\vec{k}''-\vec{q}) \\
&=& \frac{1}{2 \gamma_0 \left( 1 + \tilde{\lambda} + \tilde{\lambda}^2/2 \right)} \int \frac{d \theta''}{2 \pi}  \frac{(\openone + \cos \theta'' \sigma^y - \sin \theta'' \sigma^x)_{\mu \nu} \otimes (\openone - \cos \theta'' \sigma^y - \sin \theta'' \sigma^x)_{\lambda \kappa}}{1+ i v_\text{F} \tau_\text{e}(q_x \cos \theta'' + q_y \sin \theta'')} e^{ik\theta''} \label{eq:P_D_k_q} \label{eq:Pd} \; .
\end{eqnarray}

It is important to highlight the transposition made in the advanced Green function in order to write this new Bethe-Salpeter equation as a matrix product, which allow to drop the spin indices from now on.
As we are mainly interested in the small $\vec{q}$, we perform a Taylor expansion of Eq.~\ref{eq:Pd} around $\vec{q} = 0$~:
\begin{equation}
P^\text{D}_{(k)}(\vec{q}) = P^\text{D}_{(k)} - iv_\text{F} \tau_\text{e} \left(q_+ P^\text{D}_{(k-1)} + q_- P^\text{D}_{(k+1)}\right) - v_\text{F}^2 \tau_\text{e}^2 \left(q_+^2 P^\text{D}_{(k-2)} + 2 q_+q_- P^\text{D}_{(k)} + q_-^2 P^\text{D}_{(k+2)}\right) \; .
\end{equation}
In the absence of spin-orbit impurities, only $b_0$ is non-zero, and the only contribution to the ladder diagram is through $P_0$. When the spin-orbit impurities are present and taken into account, only 5 $b_n$ are non-zero, and 5 $P_n$ contributes to second order in $\tilde{\lambda}$, for $n=0, \pm1 , \pm2$.

Now to solve the coupled equations system, we use the fact that the disorder correlator only shows a small number of harmonics ($b_n =0$ for $n \neq 0 , \pm 1 ,\pm2$), and the fact that the Eq.~\ref{eq:Gamma} states that $\Gamma^\text{D}_{n,m} \propto b_m$ to ensure that not all the modes will contribute. Moreover as $b_n \propto \tilde{\lambda}^{\vert n \vert}$, a series expansion in powers of $\tilde{\lambda}$ is a natural choice. We write all the quantities as $X^\text{D} = \sum_\alpha X^{(\alpha)} \tilde{\lambda}^{\alpha}$, and solve the system iteratively for every order of $\tilde{\lambda}$~:
\begin{equation}
\Gamma^{(\alpha)}_{n,m}(\vec{q}) = b^{(\alpha)}_n \delta_{n,m} + \sum_{k,\alpha_1 +\alpha_2 + \alpha_3 =\alpha}\Gamma^{(\alpha_1)}_{n,m+k}(\vec{q}) P^{(\alpha_2)}_k(\vec{q}) b^{(\alpha_3)}_m \; .
\end{equation}
To zeroth order in $\tilde{\lambda}$ (so in the absence of the spin-orbit scattering), only the $\Gamma_{0,0}$ will contribute (it corresponds to the case of Dirac fermions in the presence of scalar disorder that has already been studied many times).
Then we look at the first order in $\tilde{\lambda}$, and we calculate $\Gamma_{\pm1,\pm1}$, $\Gamma_{\pm1,0}$ and also the first order contribution to $\Gamma_{0,0}$.
We have calculated this diffuson modes up to the second order in $\tilde{\lambda}$, and obtained 15 different modes $\Gamma_{n,m}$ with $n,m = 0, \pm1 \pm2$ (25 modes minus the 10 components $\Gamma_{\pm 1, \pm 2}$ , $\Gamma_{\pm2, \pm1}$ and $\Gamma_{\pm (2, -2)}$ which are of higher order in $\tilde{\lambda}$). For example, we obtain in the up-down basis along the z-direction~:
\begin{eqnarray}
   \Gamma_{-1,0} & = & \left( \begin{array}{cccc} 0&0& 0&0  \\ A_1& 0&0 & A_1 \\  X' &0& 0&X' \\ 0&0& 0&0  \end{array} \right) 
   \\
   &=& X' \left[  \frac{\sigma^x \otimes \openone + \sigma^x \otimes \sigma^z - i \sigma^y \otimes \openone - i \sigma^y \otimes \sigma^z}{4} + \frac{\openone \otimes \sigma^x + i \openone \otimes \sigma^y - \sigma^z \otimes \sigma^x  - i \sigma^z \otimes \sigma^y}{4} \right] \nonumber \\
   && + A_1 \left[  \frac{\sigma^x \otimes \openone - \sigma^x \otimes \sigma^z + i \sigma^y \otimes \openone - i \sigma^y \otimes \sigma^z}{4} + \frac{\openone \otimes \sigma^x - i \openone \otimes \sigma^y + \sigma^z \otimes \sigma^x  - i \sigma^z \otimes \sigma^y}{4} \right],
\end{eqnarray}
with $X' =\frac{i \tilde{\lambda}}{2} \left( \frac{1}{v_\text{F}^2 \tau_\text{e} ^2 q^2} - \frac{8- \tilde{\lambda}}{16} \right)$ and $A_1 = \frac{-i \tilde{\lambda} e^{-2 it}}{8} (1 -  \tilde{\lambda} )$.

\section{Expression for the Cooperon modes and their contribution to WAL}

We can solve the Bethe-Salpeter equation of the Cooperon with the same technique, or we can use the symmetry between the diffuson and the Cooperon (the Cooperon correspond to a diffuson where the advanced Green function is time-reversed). However, we know that the main contribution to the conductivity will be given by the terms in $1/q^2$, so we keep only the modes with such terms, the other ones being negligible for diffusion at long distances. We obtain the 9 modes~:

\begin{eqnarray}
\Gamma_{(0,0) \alpha \beta, \gamma \delta}^\text{C} &=& \gamma_0 \frac{1 + \tilde{\lambda}^2/4}{v_\text{F}^2 \tau_\text{e}^2 q^2} \frac{\openone_{\alpha \beta} \otimes \openone_{\gamma \delta} - \sigma^x_{\alpha \beta} \otimes \sigma^x_{\gamma \delta} - \sigma^y_{\alpha \beta} \otimes \sigma^y_{\gamma \delta} - \sigma^z_{\alpha \beta} \otimes \sigma^z_{\gamma \delta}}{4} \; ; \\
\Gamma_{(\pm1, \mp 1)\alpha \beta, \gamma \delta}^\text{C} &=&  \frac{\tilde{\lambda}^2\gamma_0}{2 v_\text{F}^2 \tau_\text{e}^2 q^2} \frac{\sigma^x_{\alpha \beta} \otimes \sigma^x_{\gamma \delta} \pm i \sigma^y_{\alpha \beta} \otimes \sigma^x_{\gamma \delta} \pm i \sigma^x_{\alpha \beta} \otimes \sigma^y_{\gamma \delta} - \sigma^y_{\alpha \beta} \otimes \sigma^y_{\gamma \delta}}{4} \; ; \\
\Gamma_{(\pm1, 0)\alpha \beta, \gamma \delta}^\text{C} &=&  \frac{- i \tilde{\lambda} \gamma_0}{ 2 v_\text{F}^2 \tau_\text{e}^2 q^2} \left( \frac{ \mp \sigma^x_{\alpha \beta} \otimes \openone_{\gamma \delta} - \sigma^x_{\alpha \beta} \otimes \sigma^z_{\gamma \delta} - i \sigma^y_{\alpha \beta} \otimes \openone_{\gamma \delta} \mp i \sigma^y_{\alpha \beta} \otimes \sigma^z_{\gamma \delta}}{4} \right. \nonumber \; ;\\
&& \qquad \left. + \frac{i \openone_{\alpha \beta} \otimes \sigma^y_{\gamma \delta} \pm \openone_{\alpha \beta} \otimes \sigma^x_{\gamma \delta} + \sigma^z_{\alpha \beta} \otimes \sigma^x_{\gamma \delta} \pm i \sigma^z_{\alpha \beta} \otimes \sigma^y_{\gamma \delta}}{4} \right) \; ;\\
\Gamma_{(0, \pm 1) \alpha \beta,\gamma \delta}^\text{C} &=&  \frac{- i \tilde{\lambda} \gamma_0}{ 2 v_\text{F}^2 \tau_\text{e}^2 q^2} \left( - \frac{ \pm \sigma^x_{\alpha \beta} \otimes \openone_{\gamma \delta} + \sigma^x_{\alpha \beta} \otimes \sigma^z_{\gamma \delta} - i \sigma^y_{\alpha \beta} \otimes \openone_{\gamma \delta} \mp i \sigma^y_{\alpha \beta} \otimes \sigma^z_{\gamma \delta}}{4} \right. \nonumber \\
&& \qquad \left. + \frac{i \openone_{\alpha \beta} \otimes \sigma^y_{\gamma \delta} \mp \openone_{\alpha \beta} \otimes \sigma^x_{\gamma \delta} - \sigma^z_{\alpha \beta} \otimes \sigma^x_{\gamma \delta} \pm i \sigma^z_{\alpha \beta} \otimes \sigma^y_{\gamma \delta}}{4} \right)\; ; \\
\Gamma_{(\pm1, \pm 1) \alpha \beta,\gamma \delta}^\text{C} &=&  \frac{\tilde{\lambda}^2\gamma_0}{2 v_\text{F}^2 \tau_\text{e}^2 q^2} \frac{\openone_{\alpha \beta} \otimes \openone_{\gamma \delta} \pm \openone_{\alpha \beta} \otimes \sigma^z_{\gamma \delta} \pm  \sigma^z_{\alpha \beta} \otimes \openone_{\gamma \delta} + \sigma^z_{\alpha \beta} \otimes \sigma^z_{\gamma \delta}}{4} \; .
\end{eqnarray}

Each one of these 9 modes contributes to the weak anti-localization when included in the three Hikami boxes (the bare one and the two dressed ones) pictured in Fig.~2 of the main text. For each mode, these three contributions can be written in the form $\delta \sigma = w \frac{e^2}{\pi h} \ln \left( \frac{\tau_{\phi}}{\tau_\text{e}} \right)$. The following table gives the value of the weight $w$ for each of the mode, and for each of the Hikami boxes.

\begin{center}
\begin{tabular}{| c | c | c | | c |}
\hline 
Mode & Bare H.B. & Dressed H.B. ($ \times$ 2) & Total $w_{\vert n \vert + \vert m \vert }$\\
\hline
\hline
$\Gamma_{0,0}$ & $1 - 3 \tilde{\lambda} + (25/4) \tilde{\lambda}^2 $& $- (1/2) + 2 \tilde{\lambda} - (19/4) \tilde{\lambda}^2$ & $(1/2) - \tilde{\lambda} + (3/2) \tilde{\lambda}^2$ \\
\hline
$\quad \Gamma_{n,m} \quad (\vert n \vert + \vert m \vert = 1) \quad $ & $(1/2) \tilde{\lambda} - (3/2) \tilde{\lambda}^2 $& $- (1/4) \tilde{\lambda} + \tilde{\lambda}^2$ & $(1/4) \tilde{\lambda} - (1/2) \tilde{\lambda}^2$ \\
\hline 
$\quad \Gamma_{n,m} \quad (\vert n \vert + \vert m \vert = 2) \quad $ & $ (1/4) \tilde{\lambda}^2 $& $- (1/8) \tilde{\lambda}^2$ & $ (1/8) \tilde{\lambda}^2$ \\
\hline 
\hline
$Total$ &$1 -  \tilde{\lambda} + (5/4) \tilde{\lambda}^2 $& $- (1/2) +  \tilde{\lambda} - (5/4) \tilde{\lambda}^2$ & $1/2 $ \\
\hline
\end{tabular}
\end{center}

This result is plotted in Fig.~3 of the main text, and shows that it is necessary to take into accounts all the Fourier modes of the Cooperon to obtain the conductivity correction $\delta \sigma = \frac{1}{2} \frac{e^2}{\pi h} \ln \left( \frac{\tau_{\phi}}{\tau_\text{e}} \right)$ characteristic of the symplectic class.

\section{Range of validity of the HLN formula}

Fig. 4 shows the crossover from the weak localization to the weak antilocalization for quasi-2DEG with parabolic dispersion when the spin-orbit impurity concentration increases. For metals, where the ratio $\tau_\phi/\tau_e$ is very large (of the order of 1000 \cite{Montambaux:book}) this crossover occurs for a value of $\tilde{\lambda}$ small enough that a perturbative treatment is possible. However, for the parameters experimentally relevant for 3DTIs, with a smaller ratio $\tau_\phi/\tau_e$ around 10 \cite{Muhlbauer:2014,Chiu:2013,Zhao:2013},  this crossover occurs for values of $\tilde{\lambda}$ of the order of the unity, which is beyond the range of validity of the HLN derivation. 
We have plotted in Fig. 4 the crossover from WL to WAL for different values of the ratio $\tau_\phi/\tau_e$ to highlight that the formula derived by Hikami \textit{et al.} is not enough to explain the WAL correction observed experimentally in 3DTIs.

\begin{figure}
\centering
\includegraphics[width= 0.70 \textwidth]{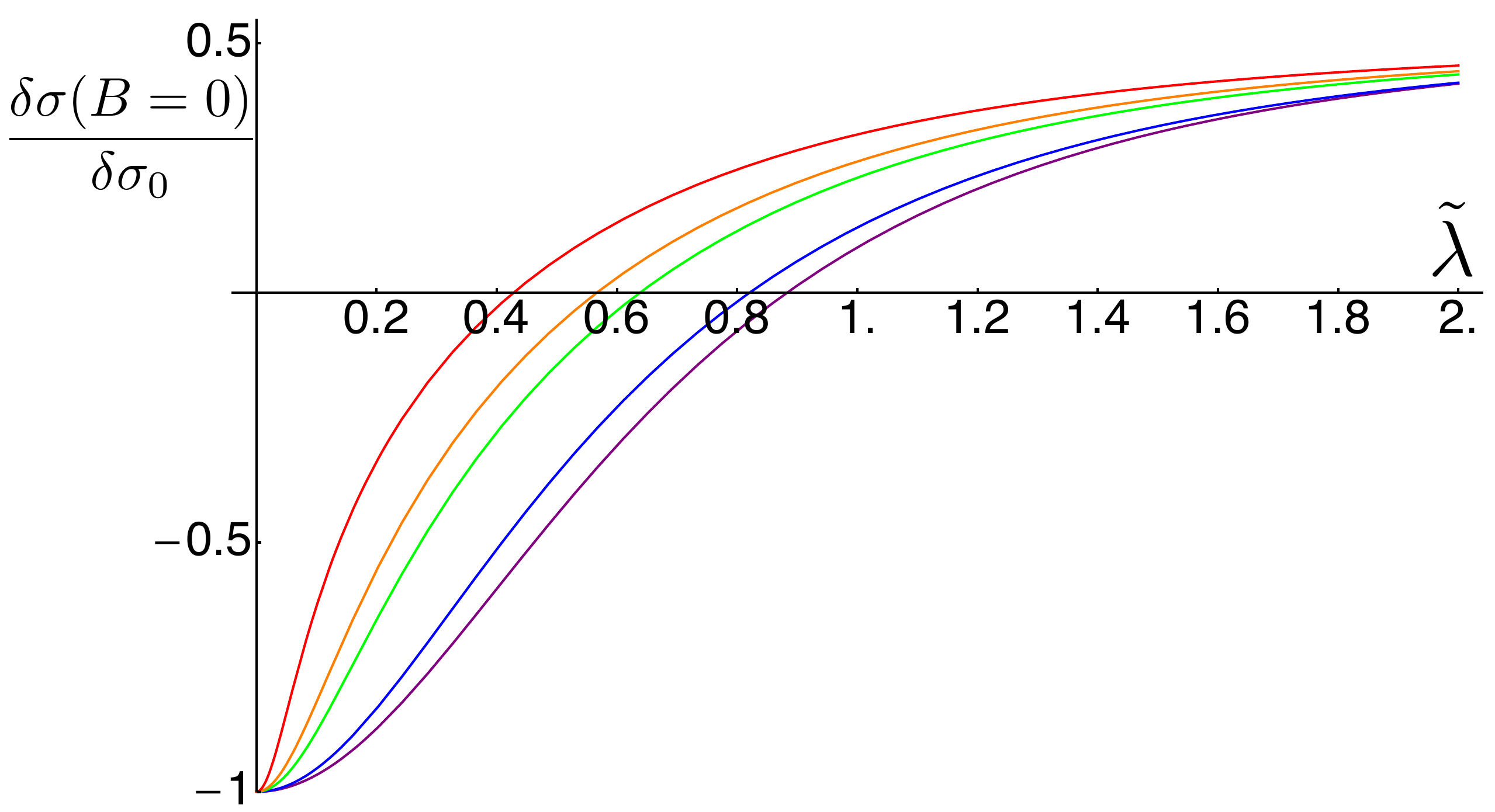}
\caption{Crossover  from WL to WAL obtained by the Hikami-Larkin-Nagaoka formula for a quasi-2DEG  for different values of the ratio $\tau_\phi / \tau_0$ over two decades from 5 (purple), 10 (blue), 50 (green), 100 (orange) to 500 (red). 
The conductivities are given in units of $\delta \sigma_0 =  \frac{e^2}{ \pi h} \ln \left( \frac{\tau_{\phi}}{\tau_\text{e}} \right)$.}
\label{fig:HLN_crossover}
\end{figure}

 \end{document}